\documentclass{article}
\usepackage[T1]{fontenc}
\usepackage[utf8]{inputenc}
\usepackage{ismir} % Remove the "submission" option for camera-ready version
\usepackage{amsmath,cite,url}
\usepackage{graphicx}
\usepackage{color}
\usepackage{graphicx} % Required for inserting images
\usepackage{url}
\usepackage{xspace}
\usepackage{pifont} % provides \ding
\usepackage{multirow}
\usepackage{multicol}
\usepackage{pifont}
\usepackage{amsmath}
\usepackage{amssymb}
\usepackage{booktabs}
\usepackage{url}
\usepackage{cite}
\usepackage{amsmath,amssymb,amsfonts}
\usepackage{algorithmic}
\usepackage{graphicx}
\usepackage{textcomp}
\usepackage{xcolor}
\usepackage{xspace}
\usepackage{booktabs}
\usepackage{multirow}
\usepackage{tikz}
\usepackage{csquotes} % automatic quote nesting using \enquote{}
\usepackage{microtype} % better typesetting
\usetikzlibrary{positioning}
\usepackage{diagbox}
\usepackage{color,colortbl} % in preamble
\usepackage{pifont}
\usepackage{tabularx}
\usepackage{adjustbox}
\usepackage{graphicx}

\title{Towards Robust Version Identification in the Wild: \\ A Dataset, Benchmark, and Fine-Tuning Study}

\oneauthor
  {Anonymous Authors}
  {Anonymous Affiliations\\\texttt{anonymous@ismir.net}}

\multauthor
  {Simon Hachmeier$^1$ \hspace{1cm} R. Oguz Araz$^2$ \hspace{1cm} Dmitry Bogdanov$^2$}
  {{\bf Robert Jäschke$^1$ \hspace{1cm} Xavier Serra$^2$}\\
  $^1$ Berlin School of Library and Information Science, Humboldt-Universität zu Berlin, Germany\\
  $^2$ Music Technology Group, Universitat Pompeu Fabra, Barcelona, Spain\\
  {\tt\small simon.hachmeier@hu-berlin.de}
  }

\def\authorname{S. Hachmeier, R. O. Araz, D. Bogdanov, R. Jäschke and X. Serra}

\begin{document}
% sizes
\newcommand{\nCsDVIdl}{95,339\xspace}
\newcommand{\nVsDVIdl}{475,699\xspace}

\newcommand{\nCsDivers}{95,260\xspace}
\newcommand{\nVsDivers}{1,102,317\xspace}

\newcommand{\nCsYVI}{44,306\xspace}
\newcommand{\nVsYVI}{629,536\xspace}

\newcommand{\cmark}{\ding{51}} % check mark
\newcommand{\xmark}{\ding{55}} % cross mark
% VI Models
\newcommand{\diversnet}{DVINetX\xspace}
\newcommand{\clews}{CLEWS\xspace}
\newcommand{\clewsl}{CLEWS$^{\text{L2}}$\xspace}
\newcommand{\ftclews}{CLEWS$^{\text{FT}}$\xspace}
\newcommand{\ftclewsl}{CLEWS$^{\text{FT+L2}}$\xspace}

\newcommand{\ftclewsma}{CLEWS$^{\text{FT+M}}$\xspace}
\newcommand{\ftclewsconfre}{CLEWS$^{\text{FT+C}}$\xspace}
\newcommand{\ftclewspenalizedconfre}{CLEWS$^{\text{FT+C+P}}$\xspace}
\newcommand{\ftclewsnoiseawarepool}{CLEWS$^{\text{FT+N}}$\xspace}

\newcommand{\clewsmini}{CLEWS-Mini\xspace}
\newcommand{\bytecoverii}{ByteCover2\xspace}
\newcommand{\bytecoveriii}{ByteCover3\xspace}
\newcommand{\bytecoveriix}{ByteCover2$^\dag$\xspace}
\newcommand{\bytecoveriiix}{ByteCover3\xspace}
\newcommand{\coverhunter}{CoverHunter\xspace}
\newcommand{\coverhunterc}{CoverHunterC\xspace}
\newcommand{\cqtnet}{CQTNet\xspace}
\newcommand{\dvinetp}{DVINet+\xspace}
\newcommand{\dvinetL}{DVINet-Large\xspace}
\newcommand{\discogsvinet}{Discogs-VINet\xspace}
\newcommand{\lyracnet}{LyraCNet\xspace}
\newcommand{\wideresnet}{WideResNet\xspace}
\newcommand{\resnet}{ResNet\xspace}

% Datasets
\newcommand{\discogsvi}{DiscogsVI\xspace}
\newcommand{\discogsviyt}{Discogs-VI-YT\xspace}
\newcommand{\dvi}{DVI\xspace}
\newcommand{\dvidl}{DVI$^*$\xspace}
\newcommand{\shsyt}{SHS-YT\xspace}
\newcommand{\shsytp}{SHS-YT$^+$\xspace}

\newcommand{\youtubevi}{YouTubeVI\xspace}
\newcommand{\yvi}{YVI\xspace} %900k
\newcommand{\yviFull}{YVI-L\xspace} %900k
\newcommand{\yviTag}{YVI-S\xspace} %200k
\newcommand{\divers}{DiVers\xspace}
\newcommand{\diversFull}{DiVers-L\xspace} %1.3M
\newcommand{\diversBalanced}{DiVers-Balanced\xspace} %1.3M
\newcommand{\diversTag}{DiVers-S\xspace} %700k
\newcommand{\shsK}{SHS100K\xspace}
\newcommand{\shsKdl}{SHS100K$^*$\xspace}

\newcommand{\datacos}{Da-TACOS\xspace}
\newcommand{\discogs}{Discogs\xspace}

% Abrreviations
\newcommand{\eg}{e.g.,\xspace}
\newcommand{\ie}{i.e.,\xspace}
\newcommand{\etc}{etc.\xspace}
\newcommand{\mir}{MIR\xspace}
\newcommand{\vi}{VI\xspace}
\newcommand{\cqt}{CQT\xspace}
\newcommand{\cqts}{CQTs\xspace}
\newcommand{\shs}{SHS\xspace}
\newcommand{\llm}{LLM\xspace}
\newcommand{\nqs}{nQs\xspace}
\newcommand{\avgcs}{avgCs\xspace}
\newcommand{\avgrs}{avgRs\xspace}

\newcommand{\qwen}{Qwen3-32B\xspace}

% Metrics
\newcommand{\map}{MAP\xspace}
\newcommand{\mr}{MR1\xspace}
\newcommand{\mrr}{MRR\xspace}
\newcommand{\nar}{NAR\xspace}

\newcommand{\ttag}[1]{\texttt{#1}}
\newcommand{\ttagset}[1]{\emph{#1}}

% tables
\newcolumntype{L}[1]{>{\raggedright\let\newline\\\arraybackslash\hspace{0pt}}p{#1}}
\newcolumntype{C}[1]{>{\centering\let\newline\\\arraybackslash\hspace{0pt}}p{#1}}
\newcolumntype{R}[1]{>{\raggedleft\let\newline\\\arraybackslash\hspace{0pt}}p{#1}}

%%%%%%%%%%%%%%%%%%%%%%  layout   %%%%%%%%%%%%%%%%%%%%%%
% max. Anzahl an Gleitobjekten auf einer Seite (oben, unten, gesamt)
\setcounter{topnumber}{5}
\setcounter{bottomnumber}{5}
\setcounter{totalnumber}{20}
\renewcommand{\topfraction}{1}
\renewcommand{\bottomfraction}{1}
\renewcommand{\textfraction}{0}
\renewcommand{\floatpagefraction}{0.99}

\maketitle

\begin{abstract}
Existing datasets for musical version identification (VI) are primarily derived from curated metadata sources such as SecondHandSongs and Discogs, and are therefore dominated by professionally recorded tracks. This leads to a domain mismatch with real-world scenarios, where amateur and user-generated content is prevalent.

To address this limitation, we introduce DiVers, a large-scale VI dataset comprising over 1.1 million musical versions, with train–validation–test splits compatible with established datasets such as Discogs-VI-YT, SHS100K, and Da-TACOS. In addition to standard version-level annotations, DiVers provides automatically assigned tags (\eg instrumental, live) and segment-level predictions indicating the presence or absence of music.

We evaluate the proposed dataset by training state-of-the-art VI systems. Our results show that models trained on DiVers achieve substantially improved robustness to acoustically diverse and noisy inputs, while maintaining a stable performance on cleaner, studio-quality benchmarks. We release the dataset metadata, code for its construction, and all experimental pipelines to support reproducibility.
\end{abstract}

\section{Introduction}
\label{sec:intro}

Musical versions are different renditions of a musical work. Automatically identifying the versions of a musical work in a set of tracks is known as version identification (\vi). While prior approaches commonly used hand-crafted audio descriptors as inputs~\cite{abrassart2022whatif,doras2020prototypical,serra2020less}, recent advances in \vi predominantly rely on learning compact representations directly from the constant-Q transform (\cqt)~\cite{araz2024mirex,du2023bytecover3,serra2025supervised}. To support the large-scale data requirements of these methods, most datasets are derived from manually curated online collections, notably SecondHandSongs (\shs)\footnote{\url{https://secondhandsongs.com/}} and Discogs.\footnote{\url{https://www.discogs.com}} %These resources correspond to what we refer to as \emph{cataloged} versions, as they are explicitly listed in structured music databases.

One example is Discogs-VI-YT (\dvi) \cite{araz2024discogs}, which contains nearly 500,000 versions and has been used to train and evaluate state-of-the-art \vi systems \cite{araz2024mirex,serra2025supervised,affolter2026scalable}. A limitation of \dvi is that it is restricted to official YouTube uploads, a design choice intended to ensure dataset cleanliness. However, this excludes user-generated or non-official uploads, which are prevalent in practice. Such \emph{in-the-wild} versions introduce additional challenges for large-scale \vi, including variations in audio quality (\eg amateur covers) and environmental noise (\eg user-recorded live performances). Moreover, music can be presented in other video content types than performances, which are associated with interruptions and non-musical content like commentary (\eg \emph{reaction videos} \cite{mcdaniel2021popular}, \emph{tutorials} \cite{airoldi2016follow}). Such variations in the audio signal are common on platforms like YouTube and can lead to notable performance degradations in \vi systems. As shown by a recent study \cite{hachmeier2025robustness}, \vi systems trained on the \shs-based dataset \shsK \cite{yu2020learning} underperform on various versions found on YouTube. Although the dataset used in this study addresses some limitations of \dvi and \shsK, its relatively small size restricts its usefulness to train \vi systems. The same limitation applies to other datasets not based on Discogs or \shs \cite{youtube_covers,ellis2007identifyingcover}. 

To address this issue, we introduce the Diverse Versions (\divers) dataset, built on the musical works of \dvi and extended with additional YouTube videos that are not required to be cataloged in curated databases. The dataset is enriched with automatically assigned tags (\eg \ttag{instrumental}, \ttag{live}) and segment-level predictions indicating the presence or absence of music. We validate its effectiveness through a benchmark study in which we train and fine-tune \vi systems. Additionally, we analyze the conditions under which fine-tuning the system \clews \cite{serra2025supervised} improves or degrades performance. Finally, we examine the embedding space shift of \clews before and after fine-tuning to better understand its effects.

To the best of our knowledge, the resulting dataset -- comprising over 1.1 million tracks with rich annotations -- is the largest and most diverse \vi dataset to date. We publicly release the dataset metadata, along with code for dataset construction\footnote{\url{https://github.com/progsi/divers_dataset}} and experiments.\footnote{\url{https://github.com/progsi/divers_benchmark}} Extracted audio features are available for research purposes upon request.%, and we provide an online application for exploration.\footnote{\url{https://divers.streamlit.app/}}

% Contributions
% - new largest and most diverse VI dataset
% - benchmark: state-of-the-art in YVI-domain and SHS-YT, slightly worse in DVI domain 
% - stratified experiment
%   - in which cases does our dataset help? (domains, tags, noise ratio)
%   - why do we get worse in DVI domain? Embedding space analysis
%   - what is the impact of the search strategy (e.g. full track vs. 20-second windows)

%\input{sections/related}
\section{Dataset}
\label{sec:dataset}

\begin{figure}
    \centering
    \includegraphics[alt={A dataset creation pipeline is shown. First, \dvi serves as the input dataset. Second, dataset metadata is used to perform YouTube searches and retrieve candidate videos. Third, retrieved candidates are filtered and matched to the input data using fuzzy matching. Fourth, noisy or irrelevant candidates are removed, followed by audio-based deduplication to eliminate duplicates. Finally, the resulting dataset is YouTube-VI (\yvi), which together with \dvi form the combined \divers dataset.},width=\linewidth]{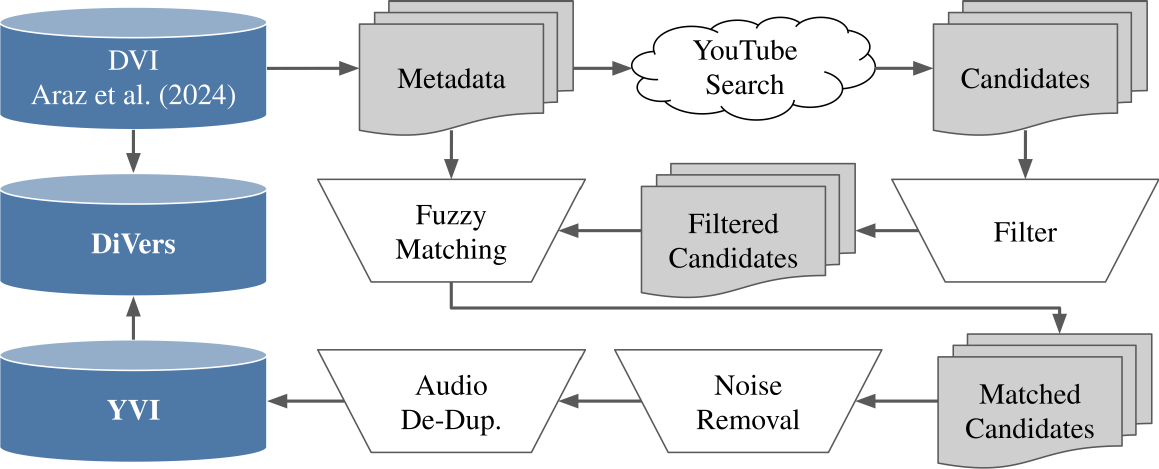}
    \caption{Dataset creation overview. Our proposed dataset \divers integrates the existing \dvi \cite{araz2024discogs} with our proposed YouTube-VI (\yvi) dataset.}
    \label{fig:dataset_creation}
\end{figure}

Figure~\ref{fig:dataset_creation} illustrates the dataset creation process. Our work builds on \dvi \cite{araz2024discogs}, which contains \emph{official} versions collected by searching YouTube using Discogs metadata (track title, writer, and performer). We do not add any new works and maintain the original partitioning of works into training, validation, and test subsets. In the following, we denote \dvidl as our version of \dvi including 96\% of versions which we could download from YouTube. For the works in this dataset, we search for new versions \emph{in-the-wild} (\eg unofficial versions such as fan-made videos).

\subsection{Finding Versions In-the-Wild}
\label{sec:creation_yvi}
%Dmitry: In general it is not clear if this method will also retrieve user-uploaded "clean" versions of the tracks, similar to the official uploads. Could this happen? How often?

Based on \dvidl, we search for version candidates on YouTube in accordance with the fair use policy \cite{youtube2025fairuse}. The search query is the normalized track title contained in the metadata of \dvidl, which is the track title in lowercase after the removal of diacritics, leading articles, punctuation marks, and text in parentheses that might indicate editing information (\eg \enquote{Radio Edit}). Per search query, we retrieve the YouTube metadata of the top 500 video search results using the \texttt{youtube\allowbreak -search\allowbreak -python} library~\cite{saini2022youtube_search}. 

Next, we exclude candidates with a duration of less than 10 seconds. This is motivated by the performance drop in \vi when using shorter segments than this duration, which might indicate a natural minimum length for the task \cite{serra2025supervised}.
Furthermore, we exclude candidates with a duration of more than 20 minutes and candidates whose YouTube identifiers are found in any of the existing datasets \dvi \cite{araz2024discogs}, \shsK \cite{yu2020learning}, and \datacos \cite{yesiler2019datacos}. That helps us to focus on finding new versions and ensuring compatibility with existing datasets. 

To promote finding \emph{unofficial} versions, we apply an additional filtering step. We exclude candidates that fulfill the requirements to be an \emph{official} version \cite{araz2024discogs}. Additionally, we exclude candidates with \emph{lyric video} or \emph{remaster} in the video title, since these cues typically indicate \emph{official} versions. After these filtering steps, we limit the set to the top 100 remaining version candidates per query. 

Next, we conduct fuzzy matching to identify new versions. To consider a candidate as a version in our dataset, we require that the work metadata from \dvidl (normalized track title, performer name) appears in the YouTube metadata (video title and description) of the candidate. From the set of all performer names that are associated with a work in \dvidl, a single match is sufficient.

We normalize performer strings in the Discogs metadata following the procedure described by \cite{hachmeier2025benchmark}. For fuzzy matching, we use the token ratio from \texttt{Rapidfuzz} \cite{bachmann2022rapidfuzz}, which was shown to be effective for this task \cite{hachmeier2024leveraging}. We consider a token ratio score of at least 80\% as a match. We exclude versions for which audio could not be downloaded from YouTube due to unavailability\footnote{Download attempts were made between November 2024 and January 2025.} or for which only a single version per work was available. 

As an additional measure to avoid retrieving merely other official versions, we perform deduplication using the \texttt{soundalike} system \cite{2026soundalike}, which builds on the \texttt{Chromaprint} framework \cite{2026chromaprint}. We apply a similarity threshold of 0.8 while retaining the default configuration for all other parameters. Deduplication is carried out at the full-track level, meaning that two items are considered duplicates only if their complete audio recordings exceed the similarity threshold. Within each set of duplicates, we prioritize retaining versions from \dvidl; otherwise, a single version is selected at random. Following this procedure, we drop 31\% of matched candidates and identify \nVsYVI newly discovered versions, denoted as \yviFull. Combined with \dvidl, this yields the \diversFull dataset, comprising a total of \nVsDivers versions.

To evaluate the quality of our dataset, we manually annotated a subset of 320 randomly selected pairs. Each pair consists of a query version (from \dvidl) and a candidate version (from \yviFull) that belong to the same work. For each pair, we assess whether the candidate was correctly assigned to the query’s work. Our analysis shows that 96.25\% of the assignments are correct. Upon investigating the errors, we found that mismatches occur when the artist is correctly identified, but the queried title matches an album title that is mentioned in the video title. In other cases, the mismatch occurred due to a wrongly assigned version in a work in \dvidl. Among the correctly matched, 6.82\% are identified as near duplicates. While we could potentially eliminate these by setting the length of the \texttt{soundalike} matching duration lower, this would potentially also exclude interesting cases (\eg reaction videos).

% errors
% same artist where  mentioned album title matches the searched title we searched for
% wrong assignments in dvi
% artist and song title are the same

%Among the correctly matched, 6.82\% are identified as near duplicates. While we could potentially eliminate these by setting the length of the matching duration lower in the deduplication, these would potentially also exclude interesting cases (\eg \ttag{reaction}). 

%https://docs.google.com/spreadsheets/d/1tMOhO6r0IF5ozxmw8sQQ3QRwLbnAQalVgmowwJ1r1UU/edit?gid=606081944#gid=606081944

\subsection{Segment-level Music/Non-Music Predictions}
\label{sec:segment_music_noise_labels}

As our dataset is expected to contain a higher proportion of non-musical content (\eg speech or crowd noise) than \dvidl, we include segment-level predictions indicating the presence of music. To obtain these, we apply PANN \cite{kong2020panns}, a sound event classification model shown to be robust in polyphonic audio settings \cite{abesser2023robust}. We run inference on sliding windows of 10~seconds with a 1-second hop. For each segment, we retain the top five predicted audio classes. Based on the sound class hierarchy analysis described in \cite{araz2024evaluation}, we define a set of music-related classes and assign a binary prediction per segment: a segment is marked as musical if any of the top five predictions correspond to a music class, and non-musical otherwise. We also use these predictions for removal of candidates containing no musical segments before deduplication of \yviFull. These account for less than 1\% of the data. The resulting segment-level predictions are retained to support downstream analyses.

\subsection{Tag Matching}
\label{sec:tag_matching}

% \begin{figure}
%     \centering
% \includegraphics[width=0.7\linewidth]{figures/diverse_tags.pdf}
%     \captionof{figure}{Schematic overview of our tagging process based on the YouTube metadata. We first create a candidate set of English tags based on two datasets and additionally defined tags. We then translate the tags into multiple languages and match these to the YouTube metadata of our dataset.}
%     \label{fig:tag_annotation}
% \end{figure}

Tags provide categorical descriptors of musical properties, such as genres (\eg \ttag{rock}, \ttag{jazz}), instruments (\eg \ttag{guitar}, \ttag{piano}), or other attributes (\eg \ttag{live}, \ttag{acoustic}, \ttag{solo}). Ideally, version representations should be invariant to such properties; however, existing \vi systems have been shown to exhibit biases with respect to them \cite{hachmeier2025robustness}. We therefore incorporate tags as auxiliary labels to support downstream analyses and stratified evaluation of representation robustness.

We construct a candidate tag set by merging two standard music auto-tagging datasets: MTG-Jamendo (MTG) \cite{bogdanov2019mtg} and MagnaTagATune (MTT) \cite{law2009evaluation}. MTG provides 195 tags covering genres, instruments, and moods/themes, with multi-word tags concatenated (\eg \ttag{heavymetal}); we restore spaces to align with MTT. MTT contributes 188 tags, including genres, instruments, and other descriptors (\eg \ttag{beat}, \ttag{solo}, \ttag{live}). After removing 68 overlapping tags, the merged set comprises 315 tags. We further add 14 YouTube-specific tags inspired by \cite{hachmeier2025robustness} and frequent $n$-grams observed in metadata (see Table~\ref{tab:tag_descriptions}), yielding 329 tags in total. While some of these tags extend beyond the traditional notion of a version (\eg \ttag{tutorial}), they capture practically relevant variations and offer a broader perspective on real-world content. To increase coverage, we translate all tags into French, German, Italian, Portuguese, and Spanish using GPT-5 \cite{singh2025openai}, followed by manual curation to ensure translation quality. We then match tags to YouTube video titles.\footnote{User-provided YouTube tags and tags detected in the description were found to be unreliable and are therefore ignored.} Matching is exact (rather than fuzzy), as translations already provide flexibility. To focus on contextual information (\eg performance characteristics), matches appearing in track titles or performer names are excluded.

\begin{table}
    \centering
    \begin{tabular}{@{}L{31.5mm}L{46mm}@{}}
    %\begin{tabularx}{8.2cm}{X X X}
    \toprule
    Tag(s) & Description \\
    \midrule
    %\ttag{backing track} & Musical track with one or more stems removed. \\
    %\midrule
    \ttag{cover} & Reinterpretation of a piece. \\
    \midrule
    \ttag{karaoke} & Track with vocals removed for singing along. \\
    %\midrule
    %\ttag{first time hearing}, \ttag{first time listening}, \ttag{reaction}, \ttag{reacts to}, \ttag{react to} & Video showing someone's response to a song. \\
    %\midrule
    %\ttag{studio} & Recording made in a controlled studio environment. \\
    \midrule
    \ttag{how to play}, \ttag{how to sing},
    \ttag{lesson}, \ttag{tutorial} & Instructional video teaching how to play or sing a song. \\
    %\midrule
    %\ttag{official} & Authorized or original release from the performer or label. \\
    \bottomrule
    \end{tabular}
    \caption{Examples of newly defined tags.}
    \label{tab:tag_descriptions}
\end{table}

We evaluate tag quality by manually annotating 30 videos for each of the 15 most frequent tags, resulting in a total of 450 annotated videos. Since the annotation considers only automatically assigned tags, this protocol estimates precision but not recall. Each video was independently annotated by two annotators, resulting in a Cohen’s $\kappa = 0.75$. Based on these annotations, the precision of automatically assigned tags exceeds 80\% for nearly all evaluated tags, with the exceptions of \ttag{rock}, \ttag{blues}, and \ttag{solo}. For \ttag{solo}, further inspection revealed that four videos contain solo sections accompanied by other instruments. While these are not purely solo performances, the highlighted passages nevertheless justify the assigned tag, illustrating the inherent ambiguity of this category and suggesting that future work could benefit from finer distinctions.

Given this encouraging precision, we use the automatically assigned tags as silver-label proxies for contextual information in our evaluation, while acknowledging that some uncertainty remains due to the automatic tagging process. Specifically, we define \yviTag as the subset of \yviFull containing videos whose titles include at least one tag, and \diversTag as the union of \dvidl and \yviTag. These subsets provide smaller alternatives to the full datasets while enabling analyses across different contextual categories.

% Tag Curation
% too general tabs: love (love in lyrics?), bass (bass instrument vs. synthetic bass; bass is almost always there), official (somewhat too fuzzy), solo (often refers to subsegments, but not necessarily solos)
% badly working tags: backing track (e.g. playalong), reaction works except for "prima volta"
% fields: video title (highest precision), description (middle) and tags (least precision)

\subsection{Analysis}
\label{sec:dataset_analysis}

 \begin{table}
 \setlength{\tabcolsep}{4pt} % default is 6pt
     \centering
    %\begin{tabular}{@{}llrrrrr@{}}
    \begin{tabularx}{8.2cm}{@{}l X *{5}{r}@{}}
    \toprule
    & & & & \multicolumn{3}{c}{Versions per work} \\
    Dataset &  Subset & Works & Versions &  Max. & Mean & Med. \\
    %\midrule
    % \multirow{3}{*}{\dvi \cite{araz2024discogs}} & Train & 80,017 & 339,771 &  455 & 4 & 2 \\
    %  &  Valid & 8,890 & 37,081 & 258 & 4 & 2 \\
    %  &  Test & 9,878 & 116,197 & 658 & 12 & 3 \\
    % \midrule
    % \multirow{3}{*}{\dvidl}& Train & 76,899 & 326,433 &  443 & 4 & 2 \\
    % &  Valid & 8,548 & 35,616 & 248 & 4 & 2 \\
    % &  Test & 9,679 & 112,790 & 637 & 12 & 3 \\
    % %\midrule
    % \multirow{3}{*}{\dvidl with 1 minimum}& Train & 78,677 & 328,694 &  445 & 4 & 2 \\
    %  &  Valid & 8,741 &  35,857 & 249 & 4 & 2 \\
    % &  Test & 9,799 & 113,021 & 637 & 12 & 3 \\
    \midrule
    \multirow{3}{*}{\textbf{\yviTag}} &  Train & 17,319 & 158,485  & 61  & 9 & 5   \\
                              &  Valid & 1,914 & 17,268 & 52 & 9 & 5 \\
                             &  Test & 3,015 & 32,650 & 60 & 11 & 7 \\
    \midrule
    \multirow{3}{*}{\textbf{\diversTag}} &  Train & 75,514 & 483,187  & 443  & 6 & 3  \\
                              &  Valid &  8,410 & 52,746  & 248 & 6 & 3 \\
                             &  Test & 9,572 & 143,436 & 623 & 15 & 5 \\
    \midrule
    \multirow{3}{*}{\textbf{\yviFull}} &  Train & 35,278 & 487,815 &  86 & 14 & 8 \\
                              & Valid & 3,870 & 53,567  & 82 & 14 & 8 \\
                             &  Test & 5,158 & 88,154 & 79 & 17 & 11 \\
    \midrule
    \multirow{3}{*}{\textbf{\diversFull}} &  Train & 77,019 & 814,189 &  461 & 11 & 4  \\
                              &  Valid & 8,570 & 89,241  & 259 & 10 & 4 \\
                             &  Test & 9,671 & 198,887 & 633 & 21 & 7 \\
    \bottomrule
    \end{tabularx}
    \caption{Number of works and versions per dataset. The minimum number of versions per work is 2 for all subsets.}
    \label{tab:dataset_overview}
 \end{table}

Table~\ref{tab:dataset_overview} reports the statistics of our proposed datasets. All exhibit a long-tail distribution in versions per work, reflected by the mean, median, and maximum statistics, consistent with real-world characteristics and the design goals of \dvi. The \yvi datasets contain a relatively high number of versions per work, but cover a smaller number of works than \dvi. Thus, \diversFull balances high class diversity with a large number of total versions by combining strengths of both datasets.

%Similarly, \yviTag has a higher mean number of versions per work than \dvi, while containing fewer total versions than \yviFull—approximately half in training, one-third in test, and just over half in validation—with a more pronounced reduction in number of works. Despite this, \diversTag still supports a larger set of works than \dvi while maintaining a slightly higher average number of versions per work.

\begin{figure}
    \centering
    \includegraphics[alt={A bar chart shows the 10 most frequently occurring tags in the dataset. The most common tag is “cover” with about 70k occurrences (6.37\%), followed by “live” with nearly 60k (5.16\%). The next most frequent tags drop significantly in frequency: “guitar” (~20k, 1.73\%), “karaoke” (~15k, 1.38\%), and “tutorial” and “reaction” (each around 10k). The remaining tags occur less frequently, including “acoustic,” “official,” “piano,” and “solo,” each contributing under 1\% of the dataset. Overall, the distribution is highly skewed, with the top two tags dominating the tag frequency.},width=\linewidth]{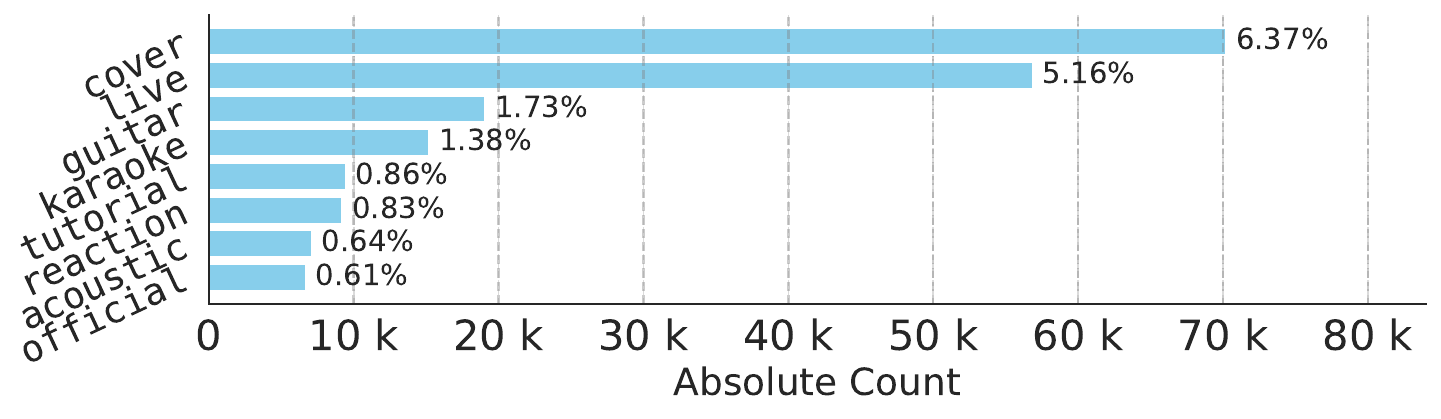}
    \caption{Most frequent matched tags in \diversFull.}
    \label{fig:matched_tags_title}
\end{figure}

% Tags
Figure~\ref{fig:matched_tags_title} shows the most frequently occurring tags after matching the original English tags and their translations. Four of our newly defined tags appear among the most common ones in video titles, namely \ttag{cover}, \ttag{karaoke}, \ttag{tutorial}, and \ttag{reaction}, highlighting the diverse contexts in which music is reinterpreted on YouTube.

% Music vs. Non-Music
Based on our sound-event–based music/non-music predictions, we find that our dataset contains 64,349~hours of music and 2,426~hours of non-music. The mean length of the longest music segment per track is 194~seconds, compared to 15~seconds for non-music segments. %Additionally, we observe that tags associated with reactions tend to have relatively low music ratios, averaging less than 60\%.

% 64,349 hours music and 2,426 hours noise --> 96% of time is music
% Mean music ratio in DVI 99.6\%, in DiVers-S 95.1\%, DiVers-L 97.8\% and in DiVers-L exclusively 96.5\%
% avg. transitions between noise and music per track 0.78
% avg. segment length music 194 seconds (music) and 15 seconds (noise)

\section{Experiments}
\label{sec:experiments}

We compare systems trained or fine-tuned on \divers against established baselines along three evaluation axes. First, we assess overall retrieval accuracy across three data regimes: \emph{Cataloged} (based on \shs and Discogs), \emph{In-the-Wild} (unconstrained web-sourced audio), and their combination (Section~\ref{sec:results_overall}). Second, we conduct stratified experiments to identify where gains are most pronounced, examining performance across tag-defined version types and non-music ratios (Section~\ref{sec:results_stratified}). Third, we analyze how fine-tuning reshapes embedding space geometry across \dvidl and \yvi (Section~\ref{sec:emb_space_analysis}).

\subsection{Training on \diversTag and \diversFull}
\label{sec:impl}

We evaluate the effectiveness of the proposed datasets for both training from scratch (Section~\ref{sec:train_diversnet}) and fine-tuning (Section~\ref{sec:ft_clews}) a state-of-the-art system. 

For each version, we extract a randomly selected 2.5-minute segment from the original audio. We compute the constant-Q transform (\cqt) using a hop size of 20~milliseconds, spanning 7 octaves starting at C1 with 12 bins per octave. The resulting representation is downsampled along the time axis by a factor of 5 and normalized to the range $[0, 1]$. Following \cite{hu2022wideresnet}, we apply SpecAugment \cite{park2019specaugment}, as well as pitch-roll and time-stretch augmentations.
Model performance is evaluated using the same combined validation metric as \cite{hu2022wideresnet}, namely the geometric mean of mean average precision (\map) and normalized average rank (\nar).

\subsubsection{Training \diversnet}
\label{sec:train_diversnet}

We adopt the architecture of \dvinetp \cite{araz2024mirex} and increase the model capacity to 20.7 million parameters by scaling the number of channels from 48 to 64 and the embedding dimension from 512 to 1,024. We refer to the resulting model as \diversnet, which we train using triplet loss with hard triplet mining and a margin of $m = 0.3$. Optimization is performed using Adam with an initial learning rate of $3 \times 10^{-4}$ and a reduce-on-plateau scheduler with a patience of 10 epochs. One epoch is defined as all versions being used as anchors. Each batch contains 24 works, with 5 versions per work. For works with fewer than 5 versions, samples are duplicated at random to meet this requirement. With these configurations, training requires 5 days on \diversTag and 7.5 days on \diversFull on a single NVIDIA H100 NVL GPU.

% Training time
% DiVers-S 5 days on single H100 NVL
% DiVers-L 7.5 days on single H100 NVL

\subsubsection{Fine-tuning \clews}
\label{sec:ft_clews}

We fine-tune pre-trained \clews on \diversTag using the best-pair-without-replacement strategy (\emph{bpwr-5}) with $k=5$, following the original proposed setup  \cite{serra2025supervised}. The method operates on non-overlapping 20-second segments extracted from the 2.5-minute track excerpts and selects the best segment pairs across pairs of versions. Training is conducted for up to 40 epochs using AdamW with a cosine learning-rate schedule, an initial learning rate of $5 \times 10^{-5}$, and weight decay of $10^{-3}$. An epoch is defined such that all versions from \dvidl are used as anchors. Due to GPU constraints, we construct batches of 25 works, each represented by 4~versions. As before, works with fewer available versions are oversampled to meet this requirement. Fine-tuning takes 4 days on two NVIDIA H100 NVL GPUs.

% Fine-Tuning time
% 4 days on two NVIDIA H100 NVL

% epoch definition to ensure cleaner pairs (anchor always clean from DVI*)

%At inference time, we apply L2 normalization to the learned representations. Although this is not part of the training procedure proposed in \cite{serra2025supervised}, we observe that this post-hoc normalization consistently improves performance across all configurations.

\subsection{Evaluation}
\label{sec:evaluation}

We evaluate our methods on our proposed test sets listed in Table~\ref{tab:dataset_overview}, as well as the test subset of \dvidl. Furthermore, we evaluate on the established \shsK dataset \cite{yu2020learning} representing another dataset in the domain of cataloged versions. We retrieve approximately 81\% of the referenced versions from YouTube, which is comparable to prior reported coverage \cite{araz2024discogs,serra2025supervised}; we denote this subset as \shsKdl. We further include \shsyt \cite{hachmeier2025robustness}, restricted to works with a minimum of two versions, forming a smaller dataset of in-the-wild versions, which we denote as \shsytp. Both additional datasets share works exclusively with the test split of \dvi, ensuring no data leakage.

% We additionally evaluate on the established \shsK test set \cite{yu2020learning}. 

We adopt the full-track evaluation protocol, representing each track with a single embedding as in common \vi benchmarks \cite{araz2024discogs,yu2020learning,yesiler2020accurate}. In our stratified experiment, we compare this global retrieval strategy against the segment-level matching approach of \cite{serra2025supervised}, which extracts 20-second segments with a 5-second hop, computes distances between segments, and reduces them into track-level distances using the \emph{bpwr-10} strategy. For both retrieval strategies, we restrict each version to its first 10 minutes to reduce computational cost. We further evaluate the effect of applying $L^2$-normalization prior to distance computation for the \clews variants. Notably, \dvinetp and \diversnet already incorporate this normalization during training. In all experiments, we report \map and \nar following \cite{serra2025supervised}, where \nar serves as a complementary metric to \map by capturing overall ranking quality beyond the top-ranked results.

In our stratified experiments, we construct controlled pairwise evaluation sets to simulate different retrieval conditions. Each subset contains $c = 500$ works, from which we sample one query version and one candidate version per work under the respective stratification constraints. In the first experiment, query versions are drawn from either \dvidl or \yvi, while candidate versions are sampled from subsets defined by specific tags, enabling analysis of performance across content-related strata. In the second experiment, queries are drawn from \dvidl and candidates are sampled from disjoint bins defined by estimated non-music ratios. To compute these bins, we estimate the ratio of non-musical content per version by averaging binary music/non-music predictions over time, obtained from PANN using 10-second segments with a 1-second hop (see Section~\ref{sec:segment_music_noise_labels}), and partition the resulting ratios into non-overlapping intervals of width 0.2. This setup yields balanced and controlled pairwise comparisons across conditions, enabling analysis of retrieval behavior under systematic variations in version type and non-music ratio.

\subsection{Results}
\label{sec:results}

\subsubsection{Overall Retrieval Accuracy}
\label{sec:results_overall}

We report the overall performance of our trained \vi systems and established baselines trained on \dvi in Table~\ref{tab:benchmark}. We observe that $L^2$ normalization of the embeddings at inference consistently improves performance for both \clews variants. \ftclewsl achieves the best performance on the \yvi datasets, \shsKdl and on \diversFull. It performs on par with \clewsl on \diversTag and \dvidl in terms of \nar, but underperforms compared to \clewsl in the \map setting on \dvidl. This points to a trade-off introduced by fine-tuning, improving robustness in noisy conditions while yielding more limited gains on cleaner datasets such as \dvidl. The contrasting behavior on \dvidl and \shsKdl may be partly explained by differences in class size: their median class sizes are 3 and 12 versions per work, respectively. Consequently, \map is more sensitive to individual retrieval failures on \dvidl.

Comparing the DVINetX variants with respect to their training data, we observe that training on \dvidl yields the best performance when evaluated on its own test subset. On all other evaluation datasets, training on DiVers-S improves NAR over training on \dvidl, whereas the corresponding MAP gains are confined to the in-the-wild subsets \yviTag and \yviFull; on \shsKdl and the combined subsets, \map remains slightly below the \dvidl-trained model. Scaling to
\diversFull improves over \diversTag on both metrics and on every evaluation dataset, and additionally surpasses training on \dvidl everywhere except on the \dvidl test subset itself. This demonstrates a clear benefit of scaling the training data. %Notably, these \diversnet variants remain competitive in terms of \nar on the \yvi datasets.

%Comparing the \diversnet variants with respect to their training data, we observe that training on \dvidl yields the best performance when evaluated on its test subset. However, for all other evaluation datasets measured in \nar, training on \diversTag consistently improves over training on \dvidl, and \diversFull further improves over \diversTag, demonstrating a clear benefit of scaling the training data. Notably, these \diversnet variants remain competitive in terms of \nar on the \yvi datasets.
 
We further compare \clewsl and \ftclewsl, as well as the \diversnet variants on \shsytp. On this dataset, \clewsl obtains a \map of 0.686 and an \nar of 10.22, while \ftclewsl improves \map to 0.708 at the cost of a slightly worse normalized average rank of 10.59. For \diversnet, its variants trained on \diversTag and \diversFull achieve \map scores of 0.636 and 0.665 respectively. Measured in \nar, the system even outperforms the \clews variants with values of 9.87 and 9.61. This is notable, because the system has only 20 million parameters compared to nearly 200 million in \clews.

\begin{table*}
\setlength{\tabcolsep}{2.5pt}
\centering
\begin{tabularx}{17.2cm}{@{}l l *{12}{>{\centering\arraybackslash}X}@{}}
\toprule
& & \multicolumn{4}{c}{Cataloged} & \multicolumn{4}{c}{In-the-Wild} & \multicolumn{4}{c}{Combined} \\
\cmidrule(lr){3-6} \cmidrule(lr){7-10} \cmidrule(lr){11-14}
Model & Train
& \multicolumn{2}{c}{\dvidl}
& \multicolumn{2}{c}{\shsKdl}
& \multicolumn{2}{c}{\yviTag}
& \multicolumn{2}{c}{\yviFull}
& \multicolumn{2}{c}{\diversTag}
& \multicolumn{2}{c}{\diversFull} \\
\cmidrule(lr){3-4}
\cmidrule(lr){5-6}
\cmidrule(lr){7-8}
\cmidrule(lr){9-10}
\cmidrule(lr){11-12}
\cmidrule(lr){13-14}
& &
\multicolumn{1}{c}{\small \map $\uparrow$} & \multicolumn{1}{c}{\small \nar $\downarrow$}
& \multicolumn{1}{c}{\small \map $\uparrow$} & \multicolumn{1}{c}{\small \nar $\downarrow$}
& \multicolumn{1}{c}{\small \map $\uparrow$} & \multicolumn{1}{c}{\small \nar $\downarrow$}
& \multicolumn{1}{c}{\small \map $\uparrow$} & \multicolumn{1}{c}{\small \nar $\downarrow$}
& \multicolumn{1}{c}{\small \map $\uparrow$} & \multicolumn{1}{c}{\small \nar $\downarrow$}
& \multicolumn{1}{c}{\small \map $\uparrow$} & \multicolumn{1}{c}{\small \nar $\downarrow$} \\
\midrule
\small \bytecoveriix \cite{du2022bytecover2} & \multirow{5}{*}{\small \dvi}
& .562 & 5.89 & .760 & 2.75 & .667 & 5.14 & .707 & 4.15 & .573 & 5.64 & .615 & 5.04 \\
\small \clews \cite{serra2025supervised} &  & \underline{.781} & 3.09 & .841 & 1.55 & .775 & 2.54  & .796 & 2.40 & .769 & 2.94 & .778 & 2.78 \\
\small \clewsl \cite{serra2025supervised} &
& \textbf{.793} & \underline{3.03} & \underline{.849} & \underline{1.37} & .802 & 1.77  & .816 & 1.91 & \underline{.785} & \underline{2.73} & \underline{.793} & \underline{2.54} \\
\small \cqtnet \cite{yu2020learning} & & .488 & 6.67 & .669 & 4.28 & .644 & 6.06 & .647 & 5.90 & .509 & 6.49 & .544 & 6.29 \\
\small \dvinetp \cite{araz2024mirex} &
& .653 & 3.74 & .775 & 2.10 & .720 & 3.40 & .724 & 3.24 & .655 & 3.63 & .670 & 3.49 \\
\midrule
\small \ftclews & \small \multirow{2}{*}{Di-S} 
& .770 & 3.24 & .844 & 1.66 & \underline{.819} & 1.99 & \underline{.824} & 1.99 & .771 & 2.94 & .784 & 2.65 \\
\small \ftclewsl & 
& \underline{.781} & \textbf{3.00} & \textbf{.857*} & \textbf{1.33*} & \textbf{.845*} & \textbf{1.31*} & \textbf{.842*} & \textbf{1.53*} & \textbf{.787} & \textbf{2.61} & \textbf{.799*} & \textbf{2.34*} \\
\midrule
\small \multirow{3}{*}{\diversnet} & \small \dvidl
& .669 & 3.67 & .791 & 1.91 & .752 & 2.44 & .756 & 2.47 & .674 & 3.37 & .693 & 3.13 \\
& \small Di-S
& .645 & 3.89 & .780 & 1.89 & .790 & 1.68 & .771 & 1.95 & .665 & 3.34 & .687 & 2.98 \\
& \small Di-L
& .654 & 3.83 & .796 & 1.67 & .804 & \underline{1.45} & .788 & \underline{1.68} & .676 & 3.23 & .701 & 2.82 \\
\bottomrule
\end{tabularx}
\caption{Results of \vi systems by training dataset across test subsets. Di abbreviates \divers and \ftclews denotes fine-tuned \clews. $^\text{L2}$ denotes $L^2$-normalized embeddings at inference time. Checkpoints for trained systems on \dvi are provided by the authors of \clews \cite{serra2025supervised}. \textbf{Bold} and \underline{underlined} values indicate the best and second-best results, respectively. * indicates a statistically significant improvement of \ftclewsl over \clewsl (Wilcoxon signed-rank test, Holm-Bonferroni-corrected $p < 0.05$).}
\label{tab:benchmark}
\end{table*}

\subsubsection{Performance Across Data Strata}
\label{sec:results_stratified}

In Table~\ref{tab:stratified_tags}, we report the difference $\Delta$ between \ftclewsl and \clewsl across stratified test subsets. Gains are observed across almost all tag-based candidate strata and both query domains, with the strongest improvements when using \yvi versions as queries. The largest and most consistent gains are obtained for the \ttag{tutorial} tag, which we attribute to its association with higher noise levels: \ttag{tutorial} content likely contains spoken segments, repeated sections, and variable pacing. In contrast, \ttag{live} shows smaller and less consistent improvements, suggesting that fine-tuning is less effective when candidate noise levels are lower. 

\begin{table}
\setlength{\tabcolsep}{3pt}
\begin{tabular}{@{}llcc@{}}
\toprule
Query & Candidate & $\Delta$ \map $\uparrow$ & $\Delta$ \nar $\downarrow$ \\
\midrule
%\multirow{10}{*}{\dvidl} & \ttag{cover} & .005\phantom{*} & -0.28* \\
\multirow{6}{*}{\dvidl} & \ttag{cover} \& \ttag{guitar} & .022* & -0.20* \\
 %& \ttag{guitar} & .051* & -0.64* \\
 & \ttag{karaoke} & .024* & -0.08* \\
 & \ttag{live} & .018* & -0.20* \\
 & \ttag{official} & -.002\phantom{*} & -0.42\phantom{*} \\
 %& \ttag{piano} & .023* & -0.12\phantom{*} \\
 %& \ttag{solo} & .043* & -0.74* \\
 & \ttag{tutorial} & .056* & -1.14* \\
 & \ttag{tutorial} \& \ttag{guitar} & .080* & -0.67* \\
 \midrule
 %\multirow{10}{*}{\yvi} & \ttag{cover} & .020* & 0.06\phantom{*} \\
\multirow{6}{*}{\yvi} & \ttag{cover} \& \ttag{guitar} & .028* & -0.12* \\
 %& \ttag{guitar} & .048* & -0.74* \\
 & \ttag{karaoke} & .027* & -0.20* \\
 & \ttag{live} & .013* & -0.20\phantom{*} \\
 & \ttag{official} & .005* & -0.36* \\
 %& \ttag{piano} & .026* & -0.34* \\
 %& \ttag{solo} & .049* & -0.71* \\
 & \ttag{tutorial} & .080* & -0.96* \\
 & \ttag{tutorial} \& \ttag{guitar} & .080* & -0.71* \\
\bottomrule
\end{tabular}
\caption{$\Delta$ between \ftclewsl and \clewsl on stratified subsets across different combinations of query domains and candidates stratified by a respective tag. * indicates statistical significance (Wilcoxon signed-rank test, Holm-Bonferroni-corrected $p < 0.05$).}
\label{tab:stratified_tags}
\end{table}

% \begin{table}
% \setlength{\tabcolsep}{3pt}
% \begin{tabular}{@{}llcc@{}}
% \toprule
% Query & Candidate & $\Delta$ \nar $\downarrow$ & $\Delta$ \map $\uparrow$ \\
% \midrule
% %\multirow{10}{*}{\dvidl} & \ttag{cover} & -0.28* & .005\phantom{*} \\
% \multirow{6}{*}{\dvidl} & \ttag{cover} \& \ttag{guitar} & -0.20* & .022* \\
%  %& \ttag{guitar} & -0.64* & .051* \\
%  & \ttag{karaoke} & -0.08* & .024* \\
%  & \ttag{live} & -0.20* & .018* \\
%  & \ttag{official} & -0.42\phantom{*} & -.002\phantom{*} \\
%  %& \ttag{piano} & -0.12\phantom{*} & .023* \\
%  %& \ttag{solo} & -0.74* & .043* \\
%  & \ttag{tutorial} & -1.14* & .056* \\
%  & \ttag{tutorial} \& \ttag{guitar} & -0.67* & .080* \\
% \midrule
%  %\multirow{10}{*}{\yvi} & \ttag{cover} & 0.06\phantom{*} & .020* \\
% \multirow{6}{*}{\yvi} & \ttag{cover} \& \ttag{guitar} & -0.12* & .028* \\
%  %& \ttag{guitar} & -0.74* & .048* \\
%  & \ttag{karaoke} & -0.20* & .027* \\
%  & \ttag{live} & -0.20\phantom{*} & .013* \\
%  & \ttag{official} & -0.36* & .005* \\
%  %& \ttag{piano} & -0.34* & .026* \\
%  %& \ttag{solo} & -0.71* & .049* \\
%  & \ttag{tutorial} & -0.96* & .080* \\
%  & \ttag{tutorial} \& \ttag{guitar} & -0.71* & .080* \\
% \bottomrule
% \end{tabular}
% \caption{$\Delta$ between \ftclewsl and \clewsl on stratified subsets across different combinations of query domains and candidates stratified to a respective tag. * indicates statistical significance (Wilcoxon signed-rank test, Holm-Bonferroni-corrected $p < 0.05$).}
% \label{tab:stratified_tags}
% \end{table}

\begin{figure}
    \centering
    \includegraphics[alt={Two heatmaps of retrieval performance under increasing amounts of non-music content in the candidate pool. The left panel shows MAP (higher is better), the right panel NAR (lower is better); columns are non-music ratios of 0.2, 0.4, 0.6 and 0.8. The four rows are grouped into full-track retrieval and segment-level matching each in combination with \clewsl and \ftclewsl. Cells are coloured on a red-to-green scale from worse to better. Both metrics degrade steadily as the non-music ratio grows, with \map falling from about 0.8 at a ratio of 0.2 to below 0.3 at 0.8, and \nar rising from below 7 to above 28. Within every column, \ftclewsl is better than \clewsl for both retrieval strategies. Segment-level matching gives the higher \map values, whereas full-track comparison gives the lower \nar values; full-track \ftclewsl is at least as good as segment-level \clewsl on both metrics at every ratio.},width=0.95\linewidth]{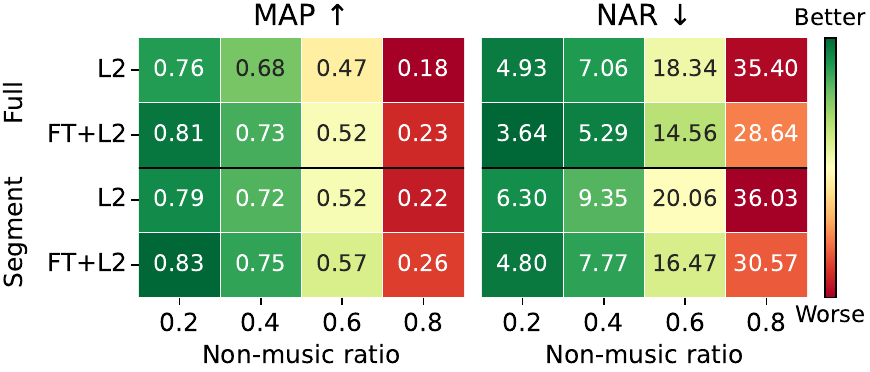}
    \caption{Retrieval performance of \clews variants compared across full track retrieval and segment-level matching with subsequent \emph{bpwr-10} reduction as proposed by \cite{serra2025supervised}.}
    \label{fig:noise_ratios}
\end{figure}

We next examine whether these gains persist across candidate non-music ratios and retrieval strategies (Figure~\ref{fig:noise_ratios}). The evaluated strategies range from efficient full-track embedding comparison to more fine-grained segment-level matching. Across both metrics, both strategies, and all non-music ratios, \ftclewsl consistently outperforms \clewsl. The margin is most evident for \nar, where the gap widens with the non-music
ratio, while the \map gains stay in a narrow band of $0.03$ to $0.05$ absolute. Comparing the two retrieval strategies, segment-level matching attains the higher \map, but is consistently worse in \nar, suggesting that local matching sharpens the top of the ranking while being more susceptible to spurious segment-level matches further down. Notably, full-track \ftclewsl matches or surpasses segment-level \clewsl on both metrics at every ratio, despite being substantially cheaper at query time. Overall, these results suggest that the improved robustness of \ftclewsl is also driven by better aggregation along the temporal axis, allowing the model to handle non-music and local distortions within a track. %At a non-music ratio of $0.8$, however, all variants degrade sharply, indicating that temporal aggregation alone does not fully compensate for heavily corrupted candidates.

Complementing the results shown in Figure~\ref{fig:noise_ratios}, we further analyze the effect of $L^2$ normalization across retrieval strategies and non-music ratios. The dominant trend is that it provides substantial gains for full-track embeddings, while offering little benefit -- and occasionally slight degradation -- for segment-level matching. For full-track embeddings, \map improves by $0.024$ – $0.048$ and \nar decreases by up to $3.13$, with gains generally increasing with non-music and being most pronounced for the fine-tuned model. In contrast, for segment-level matching, \map gains are smaller (up to $0.016$) and \nar changes remain close to zero, with degradations for the base model at higher non-music ratios ($0.022$ and $0.143$ at non-music ratios $0.6$ and $0.8$, respectively). Overall, this suggests that $L^2$ normalization improves robustness for full-track retrieval, while offering limited or occasionally negative effects for segment-level matching.

\subsubsection{Embedding Space Analysis}
\label{sec:emb_space_analysis}

We analyze how fine-tuning affects the embedding space of the \clews variants on the \dvidl and \yvi test subsets using full-track embeddings computed over 50,000 randomly sampled test versions per dataset. We measure intra-class distance as the mean pairwise distance between embeddings within the same class, and inter-class distance analogously across classes; their ratio (inter/intra) quantifies class separability. Since our best-performing configurations use $L^2$-normalized embeddings, we focus on that setting first.

Fine-tuning increases the ratio on \yvi from $3.87$ to $4.00$, but decreases it on \dvidl from $3.43$ to $3.21$, mirroring the \map degradation reported in Section~\ref{sec:results_overall}. Separability on in-the-wild versions thus improves at the cost of separability on cleaner, cataloged ones. The latter is especially consequential in \dvidl, where a median class size of 3 makes \map sensitive to the rank of individual positives.

Without normalization, fine-tuning shrinks both distances, but the intra-class distance shrinks more, so the ratio still rises on both datasets (e.g., $2.78 \rightarrow 3.23$ on \dvidl). Normalization then raises the ratio considerably for the baseline ($2.78 \rightarrow 3.43$ on \dvidl) but only marginally for the fine-tuned model ($3.23 \rightarrow 3.21$). A possible explanation is that embedding norms become more uniform across tracks after fine-tuning, with their standard deviation roughly halving on both datasets at a nearly constant mean. Normalization may therefore alter the mean-distance geometry less after fine-tuning, although it still yields consistent retrieval gains (Section~\ref{sec:results_stratified}).
\section{Conclusion and Future Work}
\label{sec:conclusion}

We introduce \divers, a large \vi dataset incorporating unofficial and user-generated YouTube recordings, resulting in a more acoustically heterogeneous benchmark than existing curated collections. Our experiments suggest that training and fine-tuning on \divers can improve robustness to noisy and diverse inputs. We further observe that fine-tuning improves class separability on in-the-wild versions while reducing it on cleaner ones from \dvidl, mirroring the corresponding \map results.

Future work should explore methods that better enforce inter-class separation and improve the robustness of learned representations. Additionally, integrating multi-view signals beyond audio embeddings -- such as lyrics \cite{affolter2026scalable,mancini2025wealy} -- offers a promising direction for more robust \vi under real-world conditions.

%Segment-level supervision signals, such as our music/non-music predictions, could be leveraged more explicitly, for instance by down-weighting or filtering non-music segments to enhance representation quality. 

%The limitations of the \emph{bpwr} strategy for selecting best pairs also warrant further investigation, for example through an adaptive $k$ fewer than a fixed number of valid matches are available. 

\section{Acknowledgments}

We thank our student assistants Sharleen Frankenstein, Siri Grenzebach, Jannis Köster and Anna Schwerdtel at the Berlin School of Library and Information Science for their support in annotating the data to evaluate the precision of the automatically matched tags.

R.~Oguz Araz is partially supported by the pre-doctoral grant AGAUR-FI Joan Oró (2024 FI-3 00065) and the Cátedra IA y Música project (TSI-100929-2023-1), funded by the Secretaría de Estado de Digitalización e Inteligencia Artificial and European Union-Next Generation EU.

% You may include an optional Acknowledgments section in your camera-ready version to refer to any individuals or organizations that should be acknowledged in your paper. \textbf{Do not include the Acknowledgments section in your submitted manuscript.} The Acknowledgments section does \textit{not} count towards the page limit for scientific content.

\section{AI Usage Statement}

We used large language models (LLMs) in a limited and supportive role during this work. Specifically, LLMs were employed for writing style refinement and language editing, as well as for assisting in code generation and subsequent refinement. In addition, LLMs were used to translate the tag vocabulary into multiple European languages to improve tag matching coverage. These uses did not constitute a methodological contribution to the research itself. All generated content, including code and translations, was manually reviewed, validated, and, where necessary, corrected by the authors to ensure accuracy and consistency. The authors take full responsibility for all aspects of the work, including the correctness of the dataset, experiments, and reported results.

\section{Ethics Statement}

This work introduces \divers, a collection of large-scale datasets of music versions collected from YouTube. All audio content referenced in the dataset is sourced from publicly accessible YouTube uploads, and no copyrighted audio is redistributed. The release contains only metadata, annotations, and derived feature representations intended for non-commercial research in music information retrieval. The dataset includes user-generated content from YouTube, which may reflect inherent demographic, cultural, and platform-driven biases. In particular, the distribution of languages, genres, and performance styles is influenced by upload behavior and search and recommendation mechanisms on the platform. As a result, models trained on \divers may inherit or amplify these biases, and we encourage careful evaluation and reporting of model behavior across different subsets of the data. The dataset is intended strictly for academic research in music information retrieval. While we do not anticipate direct harmful applications, we acknowledge that version identification technologies could, in principle, be applied in contexts such as automated copyright enforcement, which may have implications for content creators and rights holders. Researchers should therefore consider the ethical implications of downstream use cases. We release \divers to support reproducible and large-scale research in music \vi and to encourage the development of more robust and diverse retrieval systems.

% For BibTeX users:
\bibliography{bibliography}

% For non BibTeX users:
%\begin{thebibliography}{citations}
% \bibitem{Author:17}
% E.~Author and B.~Authour, ``The title of the conference paper,'' in {\em Proc.
% of the Int. Society for Music Information Retrieval Conf.}, (Suzhou, China),
% pp.~111--117, 2017.
%
% \bibitem{Someone:10}
% A.~Someone, B.~Someone, and C.~Someone, ``The title of the journal paper,''
%  {\em Journal of New Music Research}, vol.~A, pp.~111--222, September 2010.
%
% \bibitem{Person:20}
% O.~Person, {\em Title of the Book}.
% \newblock Montr\'{e}al, Canada: McGill-Queen's University Press, 2021.
%
% \bibitem{Person:09}
% F.~Person and S.~Person, ``Title of a chapter this book,'' in {\em A Book
% Containing Delightful Chapters} (A.~G. Editor, ed.), pp.~58--102, Tokyo,
% Japan: The Publisher, 2009.
%
%\end{thebibliography}

\end{document}